\documentclass[sigconf,nonacm]{acmart}

\PassOptionsToPackage{disable}{microtype}

\renewcommand\footnotetextcopyrightpermission[1]{}

\usepackage{xspace}
\newcommand{\sys}{\mbox{AgentCgroup}\xspace}

\usepackage{graphicx}

\graphicspath{{docs/img/}}

\AtBeginDocument{%
  }

\begin{document}

\title{\sys: Understanding and Controlling OS Resources of AI Agents}

% \author{Anonymous Author(s)}
% \affiliation{%
%   \institution{Anonymous Institution}
%   \city{Anonymous City}
%   \country{Anonymous Country}
% }
% \email{anonymous@example.com}

\author{Yusheng Zheng}
\affiliation{\institution{UC Santa Cruz}\country{}}
\email{yzhen165@ucsc.edu}

\author{Jiakun Fan}
\affiliation{\institution{Virginia Tech}\country{}}
\email{jiakunfan@vt.edu}

\author{Quanzhi Fu}
\affiliation{\institution{Virginia Tech}\country{}}
\email{quanzhif@vt.edu}

\author{Yiwei Yang}
\affiliation{\institution{UC Santa Cruz}\country{}}
\email{yyang363@ucsc.edu}

\author{Wei Zhang}
\affiliation{\institution{UConn}\country{}}
\email{wei.zhang@uconn.edu}

\author{Andi Quinn}
\affiliation{\institution{UC Santa Cruz}\country{}}
\email{aquinn1@ucsc.edu}

\begin{abstract}
	AI agents are increasingly deployed in multi-tenant cloud environments, where they execute diverse tool calls within sandboxed containers, each call with distinct resource demands and rapid fluctuations. We present a systematic characterization of OS-level resource dynamics in sandboxed AI coding agents, analyzing 144 software engineering tasks from the SWE-rebench benchmark across two LLM models. Our measurements reveal that (1)~OS-level execution (tool calls and container/agent initialization) accounts for 55--60\% of end-to-end task latency; (2)~memory, not CPU, is the concurrency bottleneck; (3)~memory spikes are tool-call-driven with a 15.4$\times$ peak-to-average ratio; and (4)~resource demands are highly unpredictable across tasks, runs, and models. Comparing these characteristics against serverless, microservice, and batch workloads, we identify three mismatches in existing resource controls: a granularity mismatch (container-level policies vs.\ tool-call-level dynamics), a responsiveness mismatch (user-space reaction vs.\ sub-second unpredictable bursts), and an adaptability mismatch (history-based prediction vs.\ non-deterministic stateful execution). We propose \sys, an intent-driven eBPF-based resource controller that exploits agents' ability to declare resource needs and reconstruct execution strategies, using hierarchical cgroup structures aligned with tool-call boundaries, in-kernel enforcement via sched\_ext and memcg\_bpf\_ops, and runtime-adaptive policies. Preliminary evaluation demonstrates improved multi-tenant isolation and reduced resource waste. \sys is open-source at \url{https://github.com/eunomia-bpf/agentcgroup}.
\end{abstract}

% \begin{CCSXML}
% <ccs2012>
%    <concept>
%        <concept_id>10011007.10011006.10011008.10011009.10011015</concept_id>
%        <concept_desc>Software and its engineering~Operating systems</concept_desc>
%        <concept_significance>500</concept_significance>
%    </concept>
%    <concept>
%        <concept_id>10011007.10011006.10011008.10011024</concept_id>
%        <concept_desc>Software and its engineering~Language features</concept_desc>
%        <concept_significance>300</concept_significance>
%    </concept>
% </ccs2012>
% \end{CCSXML}

% \ccsdesc[500]{Software and its engineering~Operating systems}
% \ccsdesc[300]{Software and its engineering~Language features}

% \keywords{cgroup, eBPF, resource isolation, AI agents, operating systems}

\maketitle

\section{Introduction}

AI coding agents such as Claude Code~\cite{claude-code}, OpenHands~\cite{openhands}, and SWE-agent~\cite{swe-agent} combine large language models with autonomous tool use through a reason-then-act loop~\cite{react} to address tasks such as software engineering~\cite{wang_llm_agent_survey} and OS turing~\cite{kgent}. These agents execute tool calls (compilers, test runners, package managers) inside sandboxed containers~\cite{claude-code-sandboxing,claude-secure-deployment}. With major vendors like OpenAI~\cite{openai-codex}, GitHub~\cite{github-copilot-coding-agent}, Google~\cite{google-jules}, and Devin~\cite{cognition-devin} now offering such capabilities as commercial products and cloud providers hosting many concurrent agent instances on shared infrastructure~\cite{kim2025costdynamicreasoning,bodea2025trusted,nalar}, efficient resource management becomes critical.

Effective resource management requires understanding workload behavior, yet a systematic characterization of the OS-level resource dynamics of these agent workloads is lacking. To fill this gap, we characterize these dynamics using Claude Code~\cite{claude-code} across 144 SWE-rebench~\cite{swe-rebench,swe-bench} tasks with two LLM backends: Claude Haiku 4.5 (cloud API) and GLM-4.7-Flash (local GPU). Our analysis reveals four key findings: (1)~OS-level execution (tool calls and container/agent initialization) accounts for 55--60\% of end-to-end task latency, with LLM reasoning accounting for 40--45\%; (2)~memory, not CPU, is the primary bottleneck for multi-tenant concurrency density; (3)~memory exhibits a two-layer structure with a stable ${\sim}$185\,MB framework baseline plus tool-call-driven bursts reaching a 15.4$\times$ peak-to-average ratio; and (4)~resource demands are highly unpredictable, varying 20$\times$ across tasks and 1.8$\times$ across runs of the same task.

Comparing these characteristics against existing resource controls reveals three mismatches: a \emph{granularity mismatch}, where container-level policies waste most allocated memory or trigger OOM kills; a \emph{responsiveness mismatch}, with user-space reactions at millisecond-to-minute timescales too slow for sub-second unpredictable bursts; and an \emph{adaptability mismatch}, as history-based prediction is poorly suited to non-deterministic execution where kill-and-restart destroys accumulated LLM context.

To address these mismatches, we observe that agent workloads have a key property absent in traditional workloads: agents can understand and adapt their own resource behavior. We exploit this in \sys, an intent-driven eBPF-based resource controller that combines hierarchical cgroup v2 structures aligned with tool-call boundaries, in-kernel enforcement via sched\_ext~\cite{sched-ext} and memcg\_bpf\_ops~\cite{memcg-bpf}, and runtime-adaptive policies with graceful degradation (\S\ref{sec:design}).

\noindent Our contributions are:
\begin{itemize}
	\item \textbf{Characterization.} A systematic measurement of OS-level resource dynamics in sandboxed AI coding agents (144 tasks, 2 models), yielding the four findings above (\S\ref{sec:characterization}).
	\item \textbf{Mismatch analysis.} A quantitative comparison with serverless, microservice, and batch workloads, identifying the three mismatches above (\S\ref{sec:gap}).
	\item \textbf{System.} \sys, featuring intent-driven resource adaptation, with preliminary evaluation showing 29\% lower high-priority P95 latency under multi-tenant memory contention (\S\ref{sec:design}, \S\ref{sec:eval}).
\end{itemize}

\section{Background}

\textbf{AI Coding Agents.} Modern AI coding agents combine large language models (LLMs) with tool-use capabilities to autonomously solve software engineering tasks. These agents operate in an iterative loop: the LLM reasons about the current state and emits a structured tool-use request (specifying the tool name and arguments); the agent framework parses this request, forks a subprocess inside the sandboxed container to execute the tool (e.g., running \texttt{pytest}, editing a file, invoking a compiler), collects the result, and returns it to the LLM for the next reasoning step. Representative systems include both open-source frameworks (OpenHands~\cite{openhands}, SWE-agent~\cite{swe-agent}) and commercial products (Claude Code~\cite{claude-code}, Cursor~\cite{cursor-agent}, Apple Xcode agentic coding~\cite{apple-xcode-agentic-coding}). Each tool invocation spawns distinct processes with varying resource profiles; for example, a compiler may consume gigabytes of memory, while a simple file read uses minimal resources. This creates highly dynamic, phase-varying workloads that challenge traditional resource management approaches~\cite{nalar,ni2025astraea,li2025continuum,chen2025kairos,asgar2025heterogeneous}. Recent work has begun analyzing agent costs from an inference infrastructure perspective~\cite{kim2025costdynamicreasoning,agentsight2025} and proposing OS-level abstractions for LLM agents~\cite{mei2024aios,bodea2025trusted}, but to our knowledge none provides a systematic characterization of the OS-level resource dynamics inside agent sandboxes or designs kernel-level controls informed by such measurements.

\textbf{Linux cgroup} provides a hierarchical resource governance abstraction where the kernel organizes tasks into a tree of control groups and applies controller-specific accounting and enforcement along that hierarchy~\cite{cgroupv2}. The memory controller exposes two key boundaries: \texttt{memory\allowbreak.high} as a soft throttle point that triggers reclaim pressure without invoking the OOM killer, and \texttt{memory\allowbreak.max} as a hard limit that triggers OOM when exceeded. Cgroup~v2 also provides lifecycle controls: \texttt{cgroup\allowbreak.freeze} stops all processes in a subtree until unfrozen, \texttt{cgroup\allowbreak.kill} terminates all processes while handling concurrent forks, and \texttt{memory.oom\allowbreak.group} ensures atomic OOM termination to avoid partial failures.

\textbf{eBPF} enables Linux to address the tension between standardized interfaces and dynamic workloads by introducing programmable enforcement points, providing a safe and dynamically loadable mechanism for executing control logic inside the kernel~\cite{ebpf-verifier}. On the CPU side, \texttt{sched\_ext} allows scheduling policies to be defined by BPF programs with fail-safe reversion to default behavior on errors~\cite{sched-ext}. On the memory side, \texttt{memcg\_bpf\_ops} introduces hooks such as \texttt{get\_high\_delay\_ms} for custom throttle delays on \texttt{memory.high} breaches~\cite{memcg-bpf}. These primitives enable in-kernel enforcement with microsecond-level reaction times.

\section{Agent Workload Characterization}
\label{sec:characterization}

Designing effective resource controls requires understanding workload behavior, as prior studies of serverless~\cite{shahrad2020serverless}, microservice~\cite{cortez2017resource}, and batch~\cite{verma2015borg} workloads have shown. We characterize AI coding agent workloads along three axes, each bearing on a different aspect of resource control: the execution model, which determines the granularity at which resources vary (\S\ref{sec:characterization}.2); temporal resource dynamics, which determine how fast controls must react (\S\ref{sec:characterization}.3); and cross-task variance, which determines whether demands can be predicted (\S\ref{sec:characterization}.4).

\subsection{Experimental Setup}

All experiments run on a single machine (Intel Core Ultra 9 285K, 24 cores, 128\,GB DDR5, Ubuntu 24.04, Linux 6.15.11, cgroup v2). Each task executes in an isolated Podman container using official SWE-rebench~\cite{swe-rebench} images (2.9--17.3\,GB each) with no resource limits applied. The agent framework (Claude Code) and its dependencies are bind-mounted from the host into each container, eliminating per-container installation overhead. Initialization overhead thus comprises two phases: Podman's user-namespace ID remapping of the image's overlay layers (dominant, scaling with image size), followed by agent framework startup (Node.js process and API connection). Image pull time is excluded from all reported initialization measurements. We use two underlying models: (1)~Claude Haiku 4.5 (cloud API) and (2)~GLM-4.7-Flash (local GPU). We collect 111 tasks with GLM and 33 tasks with Haiku from SWE-rebench~\cite{swe-rebench}; the 33 Haiku tasks are a subset of the 111 GLM tasks, so all cross-model comparisons use this shared overlap. For each task, we sample CPU utilization and memory usage at 1-second intervals and record each tool call's type and timestamps. Tool execution time is measured from when the LLM emits a tool-use request to when the corresponding result is returned, encompassing framework dispatch, process execution, and result collection.

\subsection{Execution Model}

\begin{figure}[t]
	\centering
	\includegraphics[width=\columnwidth]{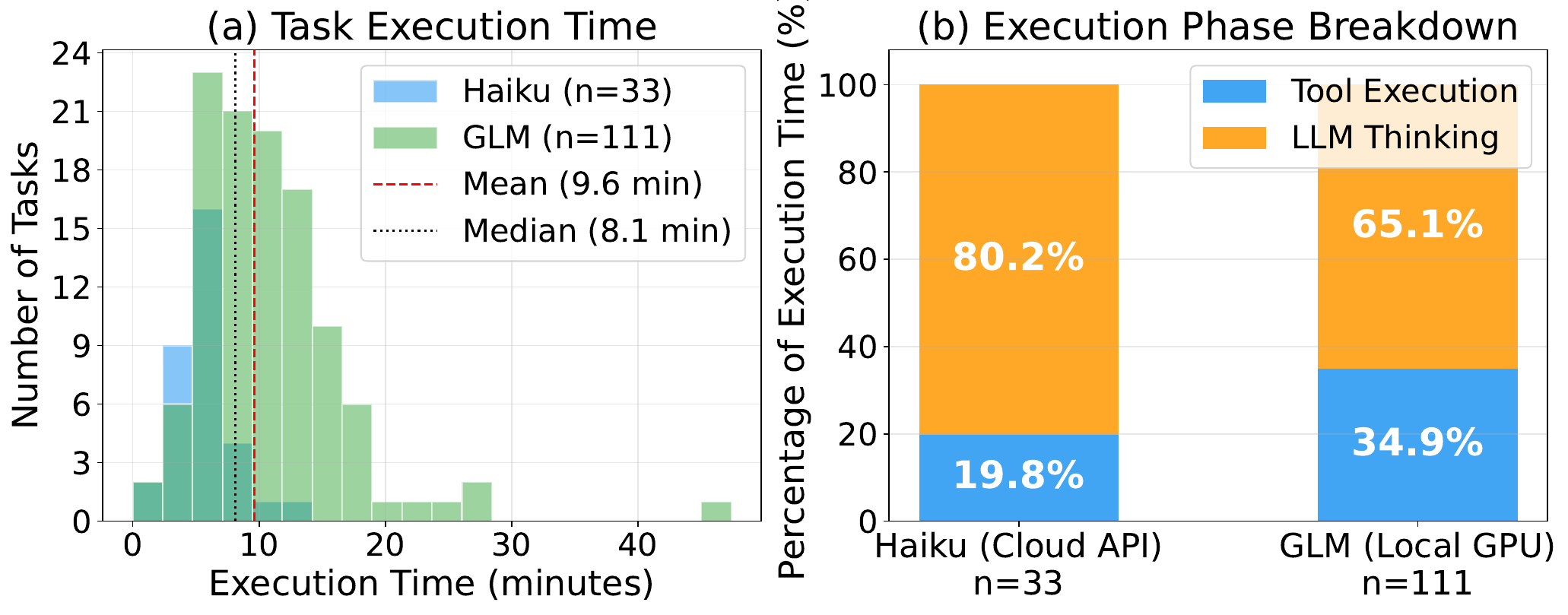}
	\caption{Task execution time distribution (a) and execution phase division (b).}
	\label{fig:exec}
\end{figure}

Each agent task runs for 5--11 minutes (GLM mean 10.8, Haiku mean 5.8, overall median 8.1; Fig.~\ref{fig:exec}(a)), far longer than serverless invocations (100\,ms--2\,s) yet shorter than batch jobs, and executes stateful multi-round reasoning and tool-call loops within a sandboxed container.

\begin{figure}[t]
	\centering
	\includegraphics[width=\columnwidth]{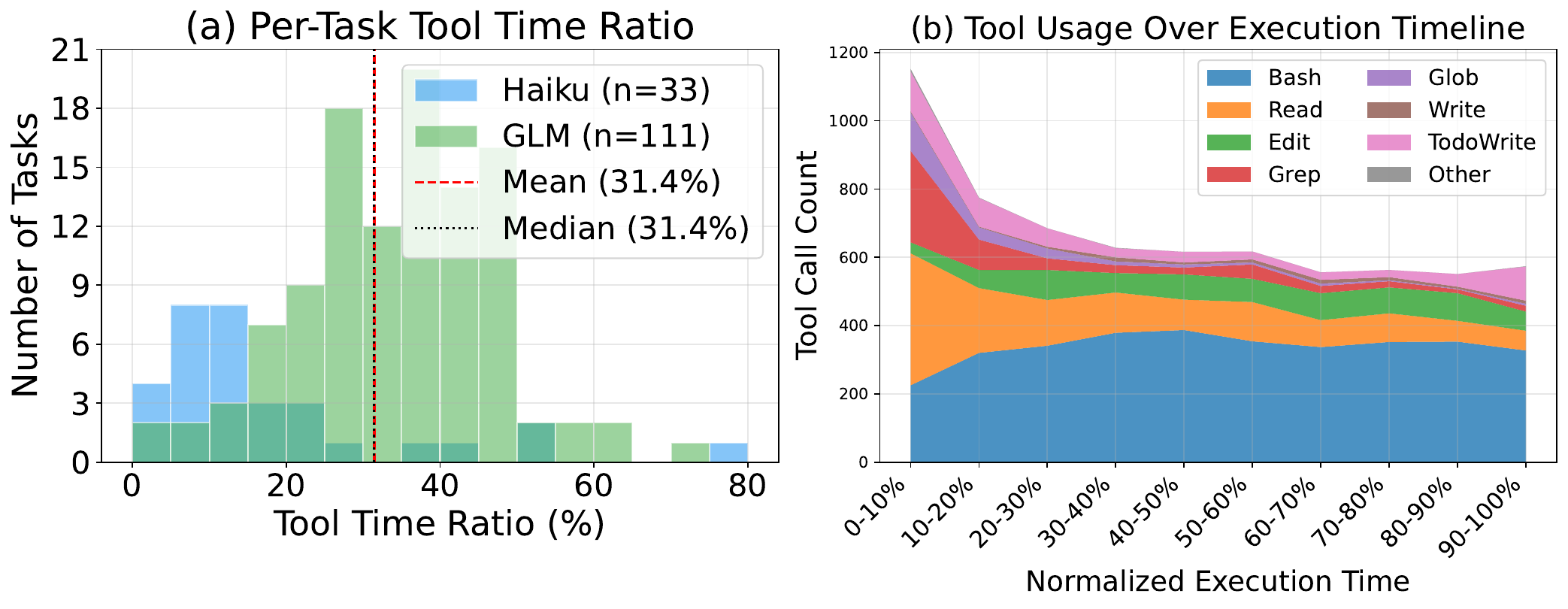}
	\caption{Tool time proportion distribution (a) and tool call distribution over execution progress (b), all 144 tasks.}
	\label{fig:tool_time}
\end{figure}

\textbf{LLM reasoning accounts for 40--45\% of end-to-end task latency; the remainder is consumed by tool execution (20--35\% of active time) and initialization (31--48\%).} As shown in Fig.~\ref{fig:exec}(b), excluding initialization overhead, tool execution accounts for a mean of 19.8\% (Haiku) and 34.9\% (GLM) of active time (median 11.6\% and 35.3\% respectively), though individual tasks range from 0\% to 86\% (Fig.~\ref{fig:tool_time}(a)). Over the full task lifecycle, container and agent initialization accounts for 31--48\%, tool execution 13--24\%, and LLM reasoning 40--45\%. OS-level overhead (initialization + tool execution) thus accounts for 55--60\% of user-perceived task completion time.

\begin{figure}[t]
	\centering
	\includegraphics[width=\columnwidth]{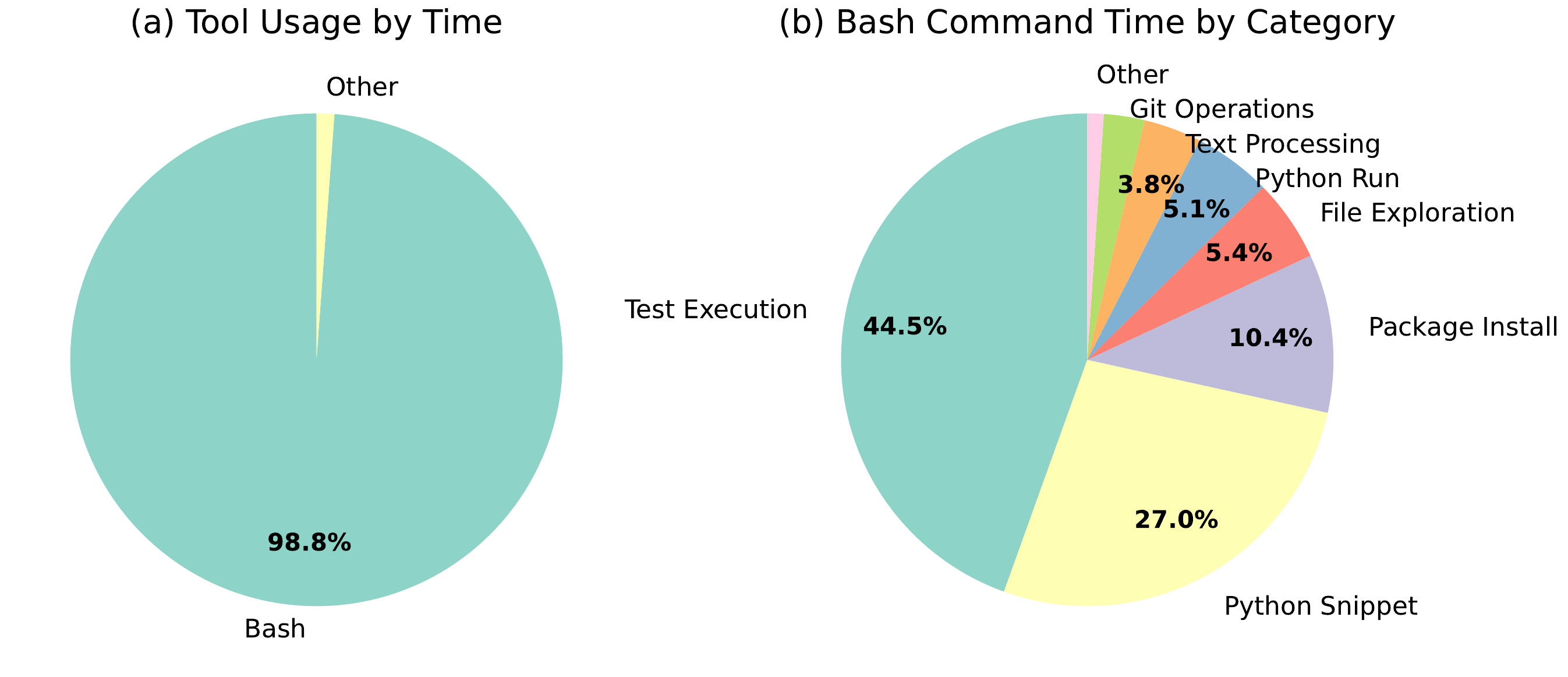}
	\caption{Tool execution time distribution (a) and Bash command semantic category proportion (b), GLM agent.}
	\label{fig:tool_type}
\end{figure}

\textbf{Bash dominates tool execution, spanning three orders of magnitude in duration.} In Haiku, Bash and the sub-agent tool (which delegates to a new agent instance) together account for over 90\% of total tool execution time (80.6\% and 17.3\% respectively); GLM relies almost entirely on Bash (98.8\% of tool time, Fig.~\ref{fig:tool_type}(a)). Execution times span three orders of magnitude: sub-agent calls average ${\sim}$18\,s, Bash commands 4--6\,s, and lightweight tools such as Read and Edit under 0.5\,s. The two models adopt different strategies: Haiku distributes work via sub-agent calls and Web search, while GLM concentrates all computation in local Bash calls.

Further analyzing Bash command semantics (Fig.~\ref{fig:tool_type}(b)), test execution (pytest, unittest, etc.) dominates Bash time in both models (Haiku 73.3\%, GLM 44.5\%), followed by package installation (${\sim}$10\%) and Python snippets (GLM 27.0\%). Resource demands vary accordingly: test execution is CPU- and memory-intensive, while file exploration and Git operations are lightweight. Tool calls also follow a temporal ``understand-modify-verify'' pattern (Fig.~\ref{fig:tool_time}(b)): Read concentrates in the first 30\% of execution, Bash peaks in 40--80\%, and Edit is distributed evenly.

\subsection{Resource Dynamics}

We now examine whether agent resource profiles match the moderate, predictable patterns assumed by existing controllers.

\textbf{Memory, not CPU, is the primary bottleneck for multi-tenant concurrency density.} Agent average CPU utilization is low (Haiku 13.2\%, GLM 7.6\%, normalized to one core, i.e., 100\%\,=\,one fully utilized core), well below saturation on the 24-core experimental platform. Peak memory can reach 2--4\,GB, so 128\,GB of RAM with peak allocation supports only 32--64 instances, while CPU utilization at this concurrency remains below 36\% of total capacity.

\begin{figure}[t]
	\centering
	\includegraphics[width=\columnwidth]{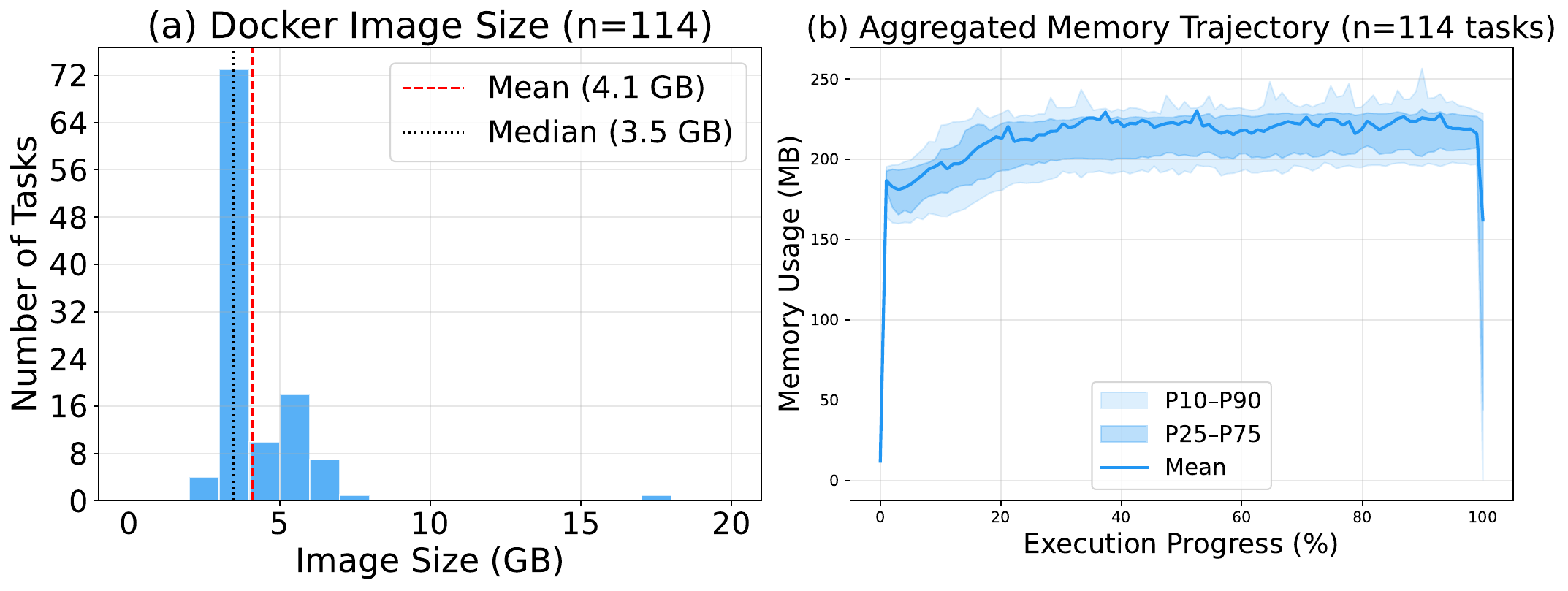}
	\caption{Docker image size distribution (a) and aggregated memory trajectory (b), all 144 tasks.}
	\label{fig:resource}
\end{figure}

\textbf{Resource consumption exhibits a two-layer structure: a ${\sim}$185\,MB framework baseline plus tool-call bursts.} As shown in Fig.~\ref{fig:resource}(b), the agent framework (e.g., Claude Code's Node.js runtime) maintains a stable memory baseline: across all 144 tasks, early-execution memory averages about 185\,MB (Haiku 183, GLM 188). Resource fluctuations come almost entirely from subprocesses spawned by tool calls: test execution, dependency installation, and other operations raise memory to 500\,MB--2\,GB, then fall back to baseline. Aggregated memory traces (Fig.~\ref{fig:resource}(b)) confirm a stable first half (${\sim}$185--200\,MB) with increasing variance in the second half as Bash-dense phases intensify. In multi-tenant scenarios, 64 concurrent instances require about 12\,GB of memory for framework baseline alone, on top of which tool bursts add an order-of-magnitude higher peak demand.

\textbf{Within the burst layer, resource consumption is determined by what the tool \emph{does} (e.g., pytest vs.\ git status), not which tool is invoked (e.g., Bash vs.\ Read): Bash calls differ by 13.7$\times$ in peak memory depending on the command executed.} For example, Bash calls in pydicom/pydicom\#2022 (a medical imaging library) consume average peak memory of 4\,GB, while streamlink/streamlink\#2160 (a network streaming tool) needs only 291\,MB. Across Bash command categories, test execution (pytest, etc.) P95 memory spike reaches 518\,MB (Haiku)/234\,MB (GLM) with average CPU spike +3.2\%; package installation P95 spike is about 233\,MB; file exploration and Git operations average only 4.5\,MB and 13.5\,MB respectively.

\begin{figure}[t]
	\centering
	\includegraphics[width=\columnwidth]{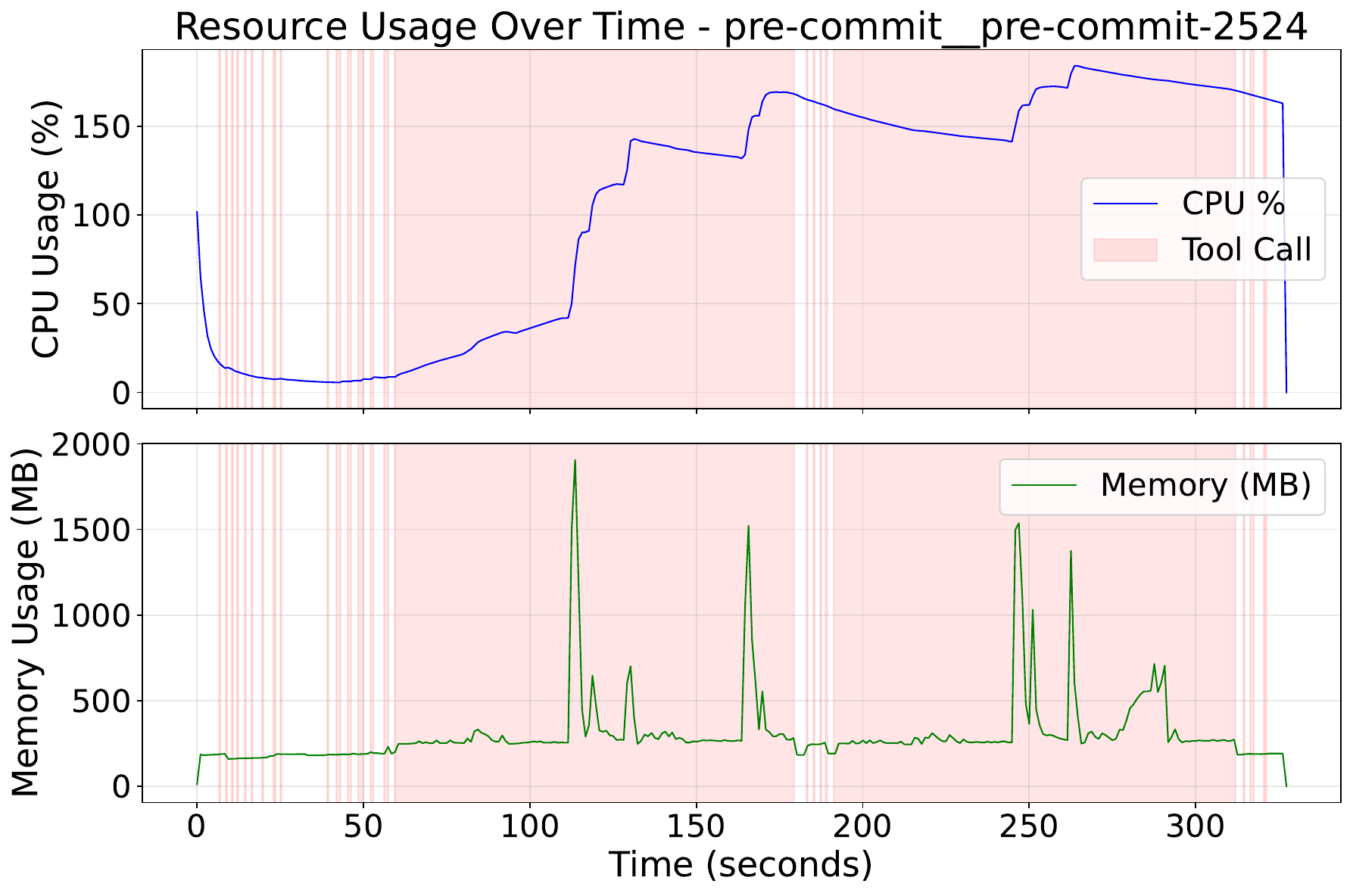}
	\caption{Resource usage time series: Haiku agent executing pre-commit/pre-commit\#2524.}
	\label{fig:timeseries_haiku}
\end{figure}

\begin{figure}[t]
	\centering
	\includegraphics[width=\columnwidth]{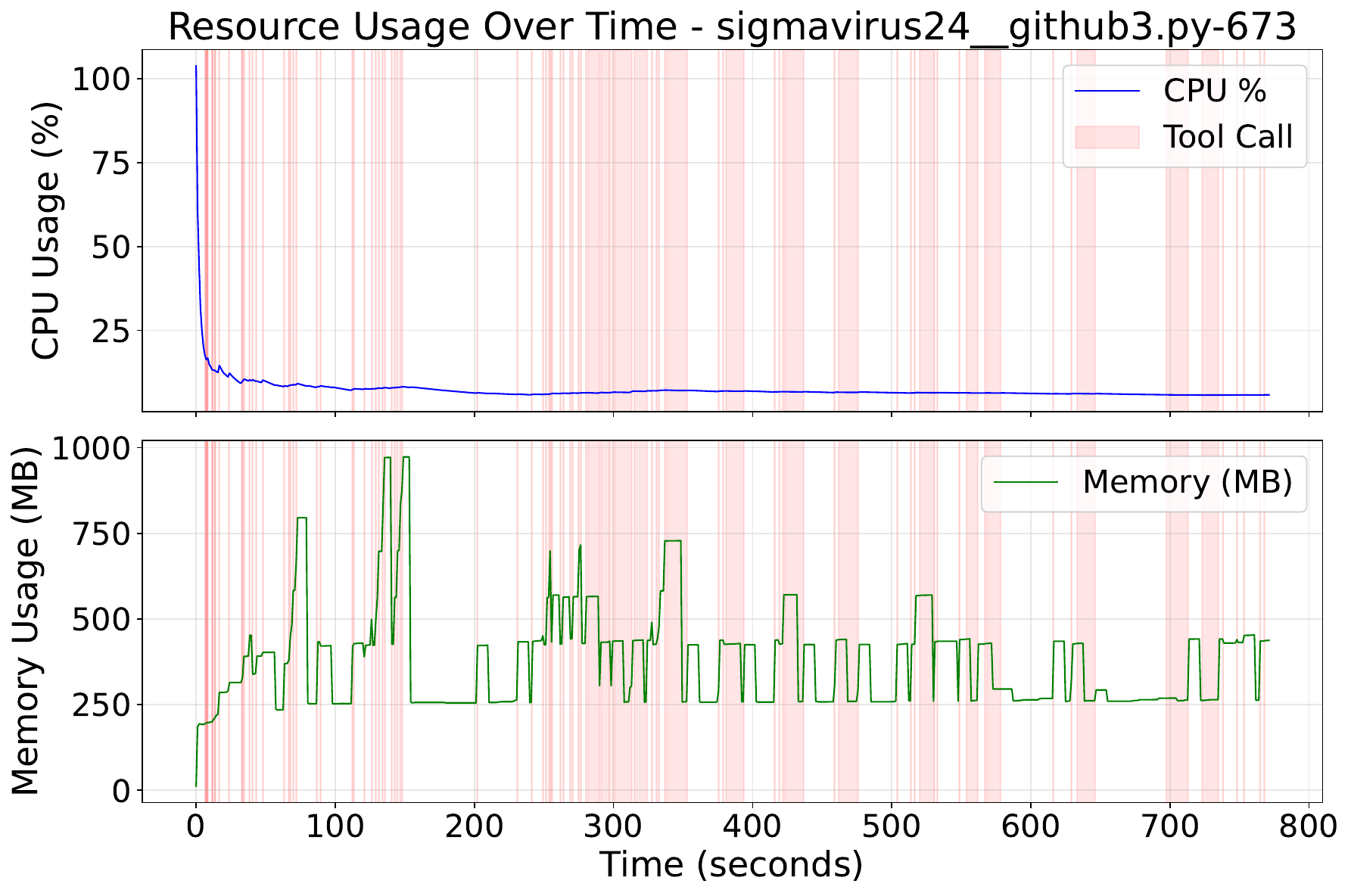}
	\caption{Resource usage time series: GLM agent executing sigmavirus24/github3.py\#673.}
	\label{fig:timeseries_glm}
\end{figure}

\textbf{Resource usage follows a burst-silence pattern, with 98.5\% of memory bursts occurring during tool calls.} Agent workloads exhibit large temporal fluctuations: memory changes by up to 2.9\,GB within a single 1-second interval and CPU peaks exceed 100\% (multi-core). As shown in Fig.~\ref{fig:timeseries_haiku} and Fig.~\ref{fig:timeseries_glm}, resource spikes align closely with tool calls, while LLM reasoning phases show stable, low usage. Classifying each 1-second sample by phase and counting memory bursts exceeding 300\,MB (${\sim}$1.6$\times$ the framework baseline): in Haiku, tool calls occupy only 28.5\% of sampling time yet contain 98.5\% of memory bursts; in GLM, 35.8\% of time contains 67.3\% of bursts (1.9$\times$ concentration). CPU burst attribution is more dispersed (Haiku 55.3\%, GLM 30.2\%), because GLM's local inference generates steady CPU load even outside tool calls.

\begin{figure}[t]
	\centering
	\includegraphics[width=\columnwidth]{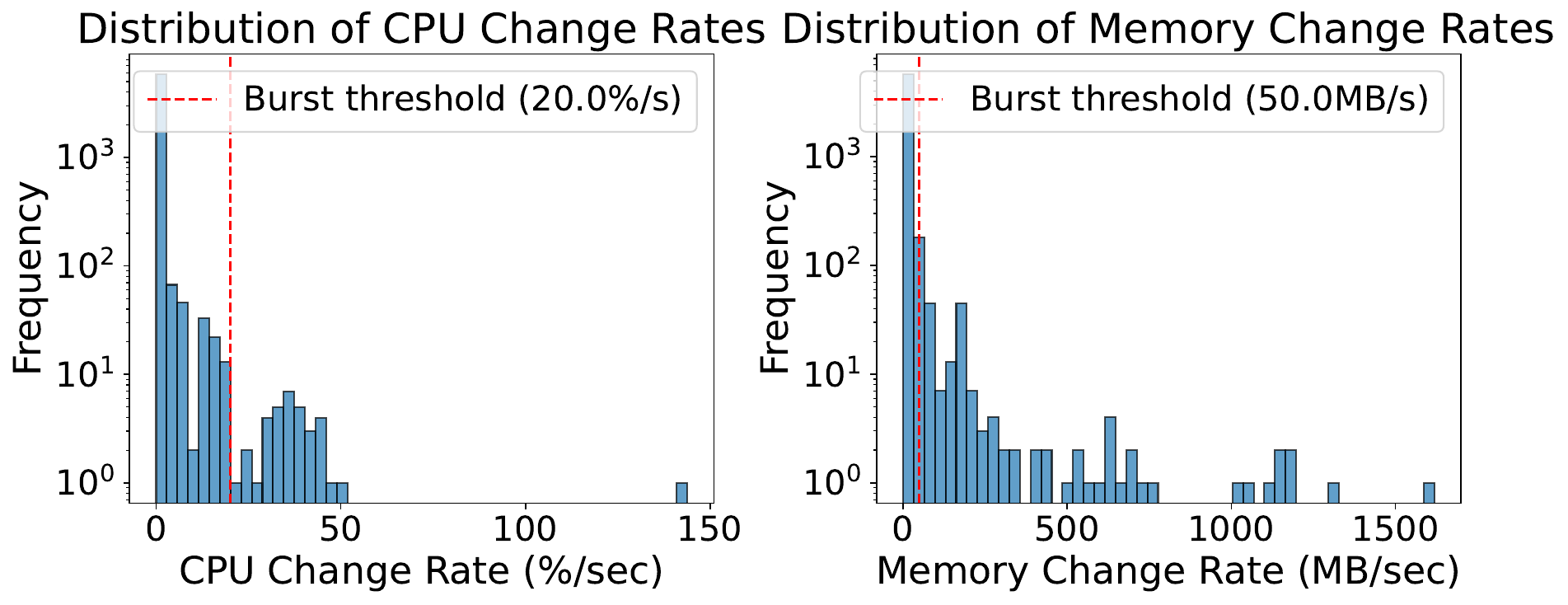}
	\caption{Resource change rate distribution (CPU and memory), Haiku dataset.}
	\label{fig:change_rate}
\end{figure}

\textbf{Resource bursts last 1--2 seconds with peak-to-average ratio up to 15.4$\times$, several times beyond traditional cloud workloads.} The most extreme case, pydicom/pydicom\#2022, has peak memory of 4060\,MB while average memory is only 264\,MB, a 15.4$\times$ ratio; this peak falls back to the 230\,MB baseline within seconds (Fig.~\ref{fig:timeseries_haiku}, Fig.~\ref{fig:timeseries_glm}). Memory peaks concentrate in the latter half of execution (mean/median $\sim$65\% progress) but occur throughout the execution cycle. Change rates are also large: maximum memory change rate reaches 3\,GB/s, CPU change rate exceeds 50\%/second, with 1.7\%--3.8\% of 1-second intervals showing memory changes exceeding 100\,MB. Defining burst thresholds at 20\%/s for CPU and 50\,MB/s for memory (dashed lines in Fig.~\ref{fig:change_rate}), a substantial tail of sampling intervals exceeds these thresholds.

\textbf{85\%--97\% of tasks contain retry loops with progressive memory accumulation.} Retry is common in agent workloads: 85\% (28/33) of tasks in the Haiku dataset contain retry groups (three or more consecutive Bash calls executing the same test command, e.g., repeated \texttt{pytest} invocations); in the GLM dataset, this ratio reaches 97\% (108/111). GLM averages 3.9 retry groups per task, with up to 56 consecutive retries, consuming 7--21\% of total execution time (Haiku 7.4\%, GLM 20.5\%). Each retry cycle retains previous memory context without cleanup, causing progressive memory accumulation (up to 502\,MB unreleased in the worst case).

\textbf{CPU-memory correlation varies by task ($-$0.84 to $+$0.50); co-directional change cannot be assumed.} The average correlation coefficient across all tasks is $-$0.39: some tasks show positive correlation (tool execution pulls up both CPU and memory simultaneously), while others show negative correlation (CPU-intensive phases coincide with lower memory demand).

\textbf{Agent container images average 3.5\,GB, 7$\times$ larger than typical microservice images and 70$\times$ larger than serverless functions.} As shown in Fig.~\ref{fig:resource}(a), image sizes concentrate in 3--4\,GB (range 2.9--17.3\,GB across 114 deduplicated images, totaling 456\,GB for 111 GLM tasks).

\subsection{Unpredictability}

\textbf{Evidence suggests high non-determinism: 1.8$\times$ execution time variance across runs of the same task.} We executed the same task (iterative/dvc\#777) three times, observing execution times of 402, 222, and 259 seconds respectively, a 1.8$\times$ difference. The three runs produced entirely different solutions: different code modifications, file counts, and strategies. This non-determinism stems from LLM reasoning randomness and decision-path diversity. Even LLM-observable proxies are uninformative: conversation rounds correlate moderately with execution time (r\,=\,+0.57 to +0.82) but not with peak memory (r\,$<$\,0.11), confirming that resource consumption is driven by what tools execute (e.g., pytest vs.\ file read) rather than reasoning scale.

\textbf{Resource demands vary 20$\times$ across tasks and diverge further across models.} Across our dataset, peak memory requirements range from 197\,MB to 4\,GB (CV\,=\,147\%): scientific computing tasks (numba/numba\#5721, pydicom/pydicom\#2022) exhibit significantly higher memory than CLI tools (joke2k/faker\#1520) or network utilities (streamlink/streamlink\#2160), yet all run in the same container. Model choice amplifies this variation: Haiku and GLM show 1.7$\times$ CPU utilization difference on the same tasks (Haiku mean 13.2\% vs.\ GLM 7.6\%). Haiku's cloud API inference consumes more local CPU for response parsing and context management; GLM's local GPU inference shifts CPU load almost entirely to tool calls (only 0.5\% of sampling points exceed 50\% CPU, vs.\ Haiku 8.2\%).

\section{Resource Management Mismatches}
\label{sec:gap}

The characterization in \S\ref{sec:characterization} reveals that agent workloads differ from known cloud workloads along several dimensions (Table~\ref{tab:comparison}). These differences create three resource management mismatches (Table~\ref{tab:mismatch}): granularity, responsiveness, and adaptability. We analyze each mismatch below, explaining why solutions from kernel cgroup interfaces to cluster-level autoscaling do not adequately address them.

\begin{table*}[t]
	\centering
	\setlength{\tabcolsep}{10pt}
	\begin{tabular}{lcccc}
		\hline
		\textbf{Dimension} & \textbf{Serverless/FaaS}~\cite{shahrad2020serverless} & \textbf{Microservices}~\cite{cortez2017resource} & \textbf{Batch/HPC}~\cite{verma2015borg} & \textbf{AI Coding Agent}                    \\
		\hline
		Execution duration & 100ms--2s                                             & Long-running                                     & Minutes--hours                          & \textbf{5--11 minutes}                      \\
		Container image    & $\sim$50\,MB                                          & 100\,MB--1\,GB                                   & 1--10\,GB                               & \textbf{2.9--17.3\,GB (med.\ 3.5)}          \\
		Statefulness       & Stateless                                             & External state                                   & Stateful                                & \textbf{In-process stateful}                \\
		Memory footprint   & 128--512\,MB                                          & Steady $\sim$1\,GB                               & Scales with data                        & \textbf{185\,MB idle, peaks 2--4\,GB}       \\
		Memory peak/avg    & $\sim$1.5$\times$                                     & 2--3$\times$                                     & $\sim$1$\times$                         & \textbf{15.4$\times$}                       \\
		CPU utilization    & Brief spike                                           & 10--40\%                                         & 80--100\%                               & \textbf{$<$13\% avg, peaks $>$175\%}        \\
		Determinism        & Deterministic                                         & Mostly deterministic                             & Deterministic                           & \textbf{1.8$\times$ variance for same task} \\
		Resource pattern   & Flat                                                  & Steady + daily cycle                             & Stable rise                             & \textbf{Burst-silence alternating}          \\
		Termination cost   & Just retry                                            & Can migrate                                      & Lose progress                           & \textbf{Lose all LLM context}               \\
		\hline
	\end{tabular}
	\caption{Quantitative comparison of AI coding agent workloads with typical cloud workloads.}
	\label{tab:comparison}
\end{table*}

\begin{table}[t]
	\centering
	\small
	\setlength{\tabcolsep}{3pt}
	\begin{tabular}{p{1.3cm}p{2.05cm}p{2.1cm}p{2.05cm}}
		\hline
		                  & \textbf{Static Limits}                                        & \textbf{Reactive Control}                                          & \textbf{Predictive Scaling}                              \\
		\hline
		\textbf{Tools}    & mem.max/high, cpu.max~\cite{cgroupv2}; K8s QoS~\cite{k8s-qos} & PSI~\cite{psi}; oomd~\cite{systemd-oomd,meta-oomd}; TMO~\cite{tmo} & VPA~\cite{k8s-vpa}; Autopilot~\cite{rzadca2020autopilot} \\
		\textbf{Assumes}  & Known peak; stable demand                                     & Gradual pressure; kill acceptable                                  & Repeatable; history valid                                \\
		\textbf{Agent}    & 15.4$\times$ peak/avg; tool-semantic variation                & 1--2\,s burst; unpredictable timing                                & 1.8$\times$ variance; kill\,=\,lose context              \\
		\textbf{Mismatch} & Granularity                                                   & Responsiveness                                                     & Adaptability                                             \\
		\hline
	\end{tabular}
	\caption{Existing resource management tools and their mismatches with agent workloads.}
	\label{tab:mismatch}
\end{table}

\subsection{Granularity Mismatch}

Agent resource demands vary at tool-call granularity, but all existing controls set a single policy at container level. The memory peak-to-average ratio illustrates this most clearly: Azure Functions exhibit near-flat memory~\cite{shahrad2020serverless}, Azure VMs stay within 2--3$\times$~\cite{cortez2017resource}, and Google Autopilot recommends within 2$\times$ of actual peaks~\cite{rzadca2020autopilot}, whereas agent workloads exhibit far higher ratios (\S\ref{sec:characterization}).

The mismatch has two dimensions: temporally, container-level policies cannot track tool-call-level dynamics; across resource types, CPU and memory are managed jointly despite being decoupled in agent workloads. On the temporal axis, cgroup~v2 hard limits (\texttt{memory\allowbreak.max}) force a binary choice: setting to peak wastes over 90\% of allocated memory (peak demand occurs less than 2\% of the time), while setting to average triggers OOM kills during tool bursts, destroying accumulated agent state. Retry loops compound this, as progressive memory accumulation means a limit adequate for early iterations may trigger OOM by the fifth (\S\ref{sec:characterization}). Soft limits (\texttt{memory\allowbreak.high}) fare no better: reclaim cannot distinguish the stable framework baseline (Node.js heap, V8 JIT cache) from tool subprocess memory, causing GC pressure on the agent runtime; moreover, a single container-level threshold cannot differentiate \texttt{git~status} (13.5\,MB) from \texttt{pytest} (518\,MB P95). Kubernetes QoS classes~\cite{k8s-qos} face both limitations: Guaranteed incurs comparable waste, BestEffort risks stateful agent termination, and Burstable still cannot set per-tool-call quotas; they also tie CPU and memory into a single class despite the weak and variable CPU--memory correlation observed in agent workloads (\S\ref{sec:characterization}). Large image sizes further preclude spawning additional containers for finer granularity (initialization consuming 31--48\% of task time), reinforcing the need for sub-container resource domains.

\subsection{Responsiveness Mismatch}

Agent resource bursts last only 1--2 seconds with change rates reaching several GB/s, a burst-silence pattern distinct from all traditional workload categories (Table~\ref{tab:comparison}). Moreover, their timing is unpredictable due to non-deterministic tool-call sequences. This combination of fast and unpredictable bursts means controllers need to react in real time at kernel speed. PSI-driven solutions (systemd-oomd~\cite{systemd-oomd}, Meta oomd~\cite{meta-oomd}) monitor cgroup Pressure Stall Information and take action (kill the cgroup or trigger swap) when pressure exceeds a threshold. Their design assumes (a) a sufficient time window for decision-making after the pressure signal appears, and (b) that kill or migration is an acceptable degradation path. Both assumptions fail for agents: the full cycle from PSI signal generation to user-space daemon reception, decision, and cgroup control file write takes tens of milliseconds, by which time the burst has already passed or triggered kernel intervention. PSI is also a container-level aggregate that cannot attribute pressure to specific tool calls; it reports aggregate container-level pressure without attribution to individual tool calls. Kubernetes VPA~\cite{k8s-vpa} adjusts resources at Pod restart level (stable) or minute-level in-place resize~\cite{k8s-inplace-resize} (alpha feature), orders of magnitude slower than second-scale tool bursts, making within-execution adjustment impossible. When slow reaction does lead to OOM, the high cost of agent container restart (\S\ref{sec:gap}) amplifies the cost of every missed burst.

\subsection{Adaptability Mismatch}

Traditional cloud workloads are largely deterministic, allowing history-based resource management: Borg's~\cite{verma2015borg} utilization data guides allocation, and Autopilot~\cite{rzadca2020autopilot} uses historical P95 metrics for automatic recommendations. Agent workloads violate this assumption along three axes. First, resource demand varies over an order of magnitude across tasks (\S\ref{sec:characterization}), so values recommended by VPA~\cite{k8s-vpa} are inevitably too high or too low. Second, even the same task produces different resource demands across runs (1.8$\times$ execution time variance), and output token count has near-zero correlation with peak memory (r\,=\,$-$0.14), making historical percentiles statistically meaningless. Third, within a single execution, retry loops (\S\ref{sec:characterization}) cause progressive memory accumulation, so memory demands grow unpredictably even during one run, not only across runs.

This unpredictability is compounded by the high cost of the traditional fallback: kill-and-restart imposes a triple penalty on agent workloads. First, slow recovery: cold-start of multi-GB containers consumes 31--48\% of total task time (\S\ref{sec:characterization}), orders of magnitude longer than serverless cold starts. Second, lost state: an OOM kill destroys minutes of accumulated in-process LLM context, and unlike microservices with external state stores, this context cannot be checkpointed or migrated. Third, non-deterministic re-execution: re-running the same task follows an entirely different solution path (\S\ref{sec:characterization}), so restart does not guarantee convergence to the original solution. Any strategy that relies on termination as fallback incurs all three penalties. Effective resource management should therefore use graceful degradation (throttling, freezing) rather than termination, and exploit agents' ability to adapt their behavior in response to resource feedback.

\section{\sys Design and Implementation}
\label{sec:design}

\sys addresses the three mismatches identified in \S\ref{sec:gap} with three corresponding mechanisms: fine-grained resource domains match tool-call-level dynamics (\emph{granularity}); in-kernel eBPF enforcement at microsecond timescales (\emph{responsiveness}); and intent-driven resource coordination replaces historical prediction with bidirectional adaptation (\emph{adaptability}).

\textbf{Fine-grained resource domains.} To address the granularity mismatch (\S\ref{sec:gap}), \sys organizes resources using a hierarchical cgroup v2 structure where each agent workload maps to a cgroup node with tool calls as child nodes, enabling per-tool-call resource constraints while maintaining overall workload budgets. Concretely, a transparent bash wrapper intercepts each \texttt{bash -c} invocation, a common tool-call execution interface in agent frameworks, and creates an ephemeral child cgroup (\texttt{tool\_<pid>\_<ts>/}) for the duration of the command. The wrapper moves the tool process into this child cgroup, collects per-tool-call resource metrics (\texttt{memory.peak}, duration), and removes the cgroup upon exit. This requires no modification to any agent framework and adds negligible overhead per invocation. \texttt{memcg\_bpf\_ops} attached to the parent cgroup automatically applies to all child cgroups via cgroup~v2 hierarchy inheritance, requiring no BPF code changes.

\textbf{In-kernel eBPF enforcement.} To address the responsiveness mismatch (\S\ref{sec:gap}), \sys executes control logic directly at kernel cgroup enforcement points via eBPF, enabling microsecond-level reaction without user-kernel round trips. On CPU, \sys uses \texttt{sched\_ext}~\cite{sched-ext} to maintain per-workload and per-tool-call metadata in BPF maps, prioritizing latency-sensitive tool calls with automatic fail-safe reversion on errors. On memory, \sys uses \texttt{memcg\_bpf\_ops} hooks~\cite{memcg-bpf} to implement custom throttling delays when a tool call cgroup breaches its soft or hard limit. When memory pressure rises, the eBPF program applies graduated enforcement based on priority: throttling via \texttt{memory.high} delays, then freezing via \texttt{cgroup.freeze}, rather than termination, preserving accumulated agent state.

\textbf{Intent-driven resource coordination.} To address the adaptability mismatch (\S\ref{sec:gap}), \sys establishes a bidirectional protocol that exploits agents' ability to adapt their resource behavior. Upward (agent $\to$ system): before each tool call, the agent may declare its expected resource need via an environment variable (e.g., \texttt{AGENT\_RESOURCE\_HINT="memory:high"} for a test suite, \texttt{"memory:low"} for a file read); the wrapper maps these to per-tool-call \texttt{memory.high} limits. Downward (system $\to$ agent): when a tool call is OOM-killed or throttled beyond recovery, the wrapper injects natural-language feedback into stderr (e.g., peak memory usage and a suggestion to reduce scope), enabling the agent to retry with a less resource-intensive approach. Declarations are advisory; the feedback loop corrects underestimates. When in-kernel enforcement alone cannot prevent a violation, this intent-driven channel provides a final graceful degradation path before termination. A lightweight user-space daemon manages cgroup lifecycle and policy configuration via shared BPF maps. We implement a proof-of-concept prototype in C using libbpf and BPF CO-RE~\cite{libbpf}, extending Agentsight~\cite{agentsight2025}. The prototype runs on Linux 6.15 with memcg\_bpf\_ops patches (currently under upstream review).

\section{Preliminary Evaluation}
\label{sec:eval}

\begin{figure}[t]
	\centering
	\includegraphics[width=\columnwidth]{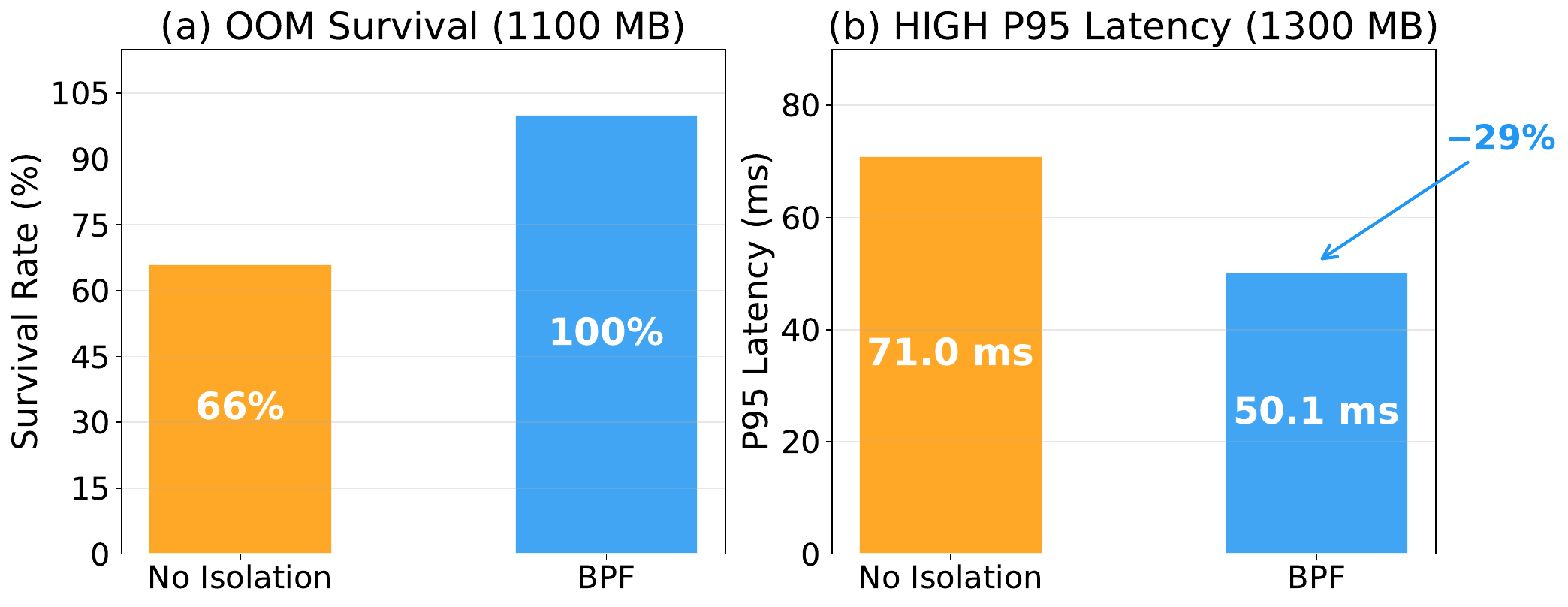}
	\caption{BPF enforcement evaluation with real agent trace replay. (a)~OOM survival rate under tight memory (1100\,MB total for ${\sim}$1233\,MB demand). (b)~HIGH-priority P95 allocation latency under moderate memory (1300\,MB).}
	\label{fig:eval}
\end{figure}

We evaluate \sys by replaying real agent memory traces from \S\ref{sec:characterization} at 50$\times$ accelerated speed in a multi-tenant setting with \sys enforcement. Experiments run on an Intel Core Ultra 7 258V (4 cores, 16\,GB RAM) with a patched Linux 6.19.0-rc5 kernel (bpf-next tree plus the memcg struct\_ops RFC patches~\cite{memcg-bpf}). No application code is modified; isolation is achieved entirely through container cgroup boundaries and eBPF hooks. We use three agent traces run concurrently in separate cgroups: dask/dask\#11628 as the HIGH-priority session (peak 421\,MB, \texttt{memory.high\,=\,max}) and two instances of sigmavirus24/github3.py\#673 as LOW-priority sessions (peak 406\,MB each, \texttt{memory.high\,=\,400\,MB}). Under BPF, LOW cgroups are throttled when the HIGH cgroup experiences memory pressure; the HIGH cgroup is protected via \texttt{below\_low}. We compare against a no-isolation baseline under two memory pressure scenarios (Fig.~\ref{fig:eval}).

\textbf{Results.} Under tight memory (1100\,MB total for ${\sim}$1233\,MB combined demand), the baseline OOM-kills one LOW process (66\% survival); BPF allows all processes to complete (100\%) by throttling LOW allocations (239 delay triggers) while HIGH finishes with only +2.8\% overhead (Fig.~\ref{fig:eval}(a)). BPF also reduces HIGH-priority P95 allocation latency by 29\% (70.97$\to$50.14\,ms) through reduced memory contention (Fig.~\ref{fig:eval}(b)). Enforcement overhead is negligible: P50 latency increases by 0.3\% and total completion time decreases by 1.1\%. Separately, kernel selftests with a 2000\,ms configured delay confirm BPF throttling precision within 2.3\% relative error (2.000\,$\pm$\,0.046\,s measured).

\section{Conclusion and Future Work}

We characterized AI coding agent tasks across two models, finding OS execution accounts for 55--60\% of latency with a memory peak-to-average ratio of 15.4$\times$. Based on these findings, we presented \sys, an intent-driven eBPF-based resource controller addressing granularity, responsiveness, and adaptability mismatches through tool-call-aligned cgroup domains, microsecond-level in-kernel reaction, and resource adaptation that exploits agents' ability to reconstruct execution strategies. Our current evaluation is limited to trace replay with a proof-of-concept prototype; the characterization covers one agent framework and one benchmark. Future work will evaluate \sys with live agent workloads at production scale, validate intent-driven adaptation across diverse tasks and agent frameworks, and explore fine-grained resource control across diverse container runtimes. Moreover, the current prototype mainly focuses on CPU and memory resource control; challenges such as initialization overhead, large container images, and retry-induced accumulation remain open and require optimizations across the sandbox runtime and orchestration stack.

\bibliographystyle{ACM-Reference-Format}
\bibliography{references}

@misc{cgroupv2,
  author = {Heo, Tejun},
  title = {Control Group v2},
  howpublished = {Linux Kernel Documentation},
  year = {2015},
  note = {\url{https://docs.kernel.org/admin-guide/cgroup-v2.html}},
}

@misc{ebpf-verifier,
  author = {{Linux Kernel Community}},
  title = {{eBPF} Verifier},
  howpublished = {Linux Kernel Documentation},
  year = {2024},
  note = {\url{https://docs.kernel.org/bpf/verifier.html}},
}

@misc{sched-ext,
  author = {Heo, Tejun and Vernet, David and Don, Josh},
  title = {Extensible Scheduler Class},
  howpublished = {Linux Kernel Documentation},
  year = {2023},
  note = {\url{https://docs.kernel.org/scheduler/sched-ext.html}},
}

@misc{memcg-bpf,
  author = {Zhu, Hui},
  title = {mm: memcontrol: Add {BPF} hooks for memory controller},
  howpublished = {LWN.net},
  year = {2026},
  note = {RFC PATCH bpf-next v3 00/12, Jan 23 2026. \url{https://lwn.net/Articles/1055698/}},
}

@misc{libbpf,
  author = {{Linux Kernel Community}},
  title = {libbpf Overview},
  howpublished = {Linux Kernel Documentation},
  year = {2024},
  note = {\url{https://docs.kernel.org/bpf/libbpf/libbpf_overview.html}},
}

@inproceedings{tmo,
  author = {Weiner, Johannes and Agarwal, Niket and Schatzberg, Dan and Yang, Leon and Wang, Hao and Sanouillet, Blaise and Sharma, Bikash and Heo, Tejun and Jain, Mayank and Tang, Chunqiang and Skarlatos, Dimitrios},
  title = {{TMO}: Transparent Memory Offloading in Datacenters},
  booktitle = {Proceedings of the 27th ACM International Conference on Architectural Support for Programming Languages and Operating Systems (ASPLOS '22)},
  year = {2022},
  pages = {609--621},
  publisher = {ACM},
  doi = {10.1145/3503222.3507731},
}

@misc{nalar,
  author = {Laju, Marco and Son, Donghyun and Agarwal, Saurabh and Kedia, Nitin and Lee, Myungjin and Srinivasa, Jayanth and Akella, Aditya},
  title = {Nalar: An Agent Serving Framework},
  howpublished = {arXiv preprint arXiv:2601.05109},
  year = {2026},
}

@misc{claude-code,
  author = {{Anthropic}},
  title = {How Claude Code Works},
  howpublished = {Claude Code Documentation},
  year = {2026},
  note = {\url{https://code.claude.com/docs/en/how-claude-code-works} (accessed 2026-02-09)},
}

@inproceedings{openhands,
  author = {Wang, Xingyao and Li, Boxuan and Song, Yufan and Xu, Frank F. and Tang, Xiangru and Zhuge, Mingchen and Pan, Jiayi and Song, Yueqi and Li, Bowen and Singh, Jaskirat and Tran, Hoang H. and Li, Fuqiang and Ma, Ren and Zheng, Mingzhang and Qian, Bill and Shao, Yanjun and Muennighoff, Niklas and Zhang, Yizhe and Hui, Binyuan and Lin, Junyang and Brennan, Robert and Peng, Hao and Ji, Heng and Neubig, Graham},
  title = {{OpenHands}: An Open Platform for {AI} Software Developers as Generalist Agents},
  booktitle = {International Conference on Learning Representations (ICLR)},
  year = {2025},
  note = {arXiv:2407.16741},
}

@inproceedings{swe-agent,
  author = {Yang, John and Jimenez, Carlos E. and Wettig, Alexander and Lieret, Kilian and Yao, Shunyu and Narasimhan, Karthik and Press, Ofir},
  title = {{SWE}-agent: Agent-Computer Interfaces Enable Automated Software Engineering},
  booktitle = {Advances in Neural Information Processing Systems (NeurIPS)},
  year = {2024},
  note = {arXiv:2405.15793},
}

@misc{swe-rebench,
  author = {Badertdinov, Ibragim and Golubev, Alexander and Nekrashevich, Maksim and Shevtsov, Anton and Karasik, Simon and Andriushchenko, Andrei and Trofimova, Maria and Litvintseva, Daria and Yangel, Boris},
  title = {{SWE-rebench}: An Automated Pipeline for Task Collection and Decontaminated Evaluation of Software Engineering Agents},
  howpublished = {arXiv preprint arXiv:2505.20411},
  year = {2025},
  note = {Dataset: \url{https://huggingface.co/datasets/nebius/SWE-rebench}},
}

@inproceedings{shahrad2020serverless,
  author = {Shahrad, Mohammad and Fonseca, Rodrigo and Goiri, \'{I}\~{n}igo and Chaudhry, Gohar and Batum, Paul and Cober, Jason and Laureano, Esdras and Trespass, Christos and Russinovich, Mark and Bianchini, Ricardo},
  title = {Serverless in the Wild: Characterizing and Optimizing the Serverless Workload at a Large Cloud Provider},
  booktitle = {Proceedings of the 2020 USENIX Annual Technical Conference (ATC)},
  year = {2020},
  pages = {205--218},
  publisher = {USENIX},
}

@inproceedings{rzadca2020autopilot,
  author = {Rzadca, Krzysztof and Findeisen, Pawel and Swiderski, Jacek and Zych, Przemyslaw and Broniek, Przemyslaw and Kusmierek, Jarek and Nowak, Pawel and Strack, Ben and Witusowski, Piotr and Hand, Steven and Wilkes, John},
  title = {Autopilot: Workload Autoscaling at {Google} Scale},
  booktitle = {Proceedings of the Fifteenth European Conference on Computer Systems (EuroSys)},
  year = {2020},
  publisher = {ACM},
}

@inproceedings{verma2015borg,
  author = {Verma, Abhishek and Pedrosa, Luis and Korupolu, Madhukar and Oppenheimer, David and Tune, Eric and Wilkes, John},
  title = {Large-Scale Cluster Management at {Google} with {Borg}},
  booktitle = {Proceedings of the Tenth European Conference on Computer Systems (EuroSys)},
  year = {2015},
  publisher = {ACM},
}

@inproceedings{cortez2017resource,
  author = {Cortez, Eli and Bonde, Anand and Muzio, Alexandre and Russinovich, Mark and Fontoura, Marcus and Bianchini, Ricardo},
  title = {Resource Central: Understanding and Predicting Workloads for Improved Resource Management in Large Cloud Platforms},
  booktitle = {Proceedings of the 26th Symposium on Operating Systems Principles (SOSP)},
  year = {2017},
  pages = {153--167},
  publisher = {ACM},
}

@misc{k8s-vpa,
  author = {{Kubernetes Community}},
  title = {Vertical Pod Autoscaler},
  howpublished = {Kubernetes Documentation},
  year = {2024},
  note = {\url{https://github.com/kubernetes/autoscaler/tree/master/vertical-pod-autoscaler}},
}

@misc{k8s-qos,
  author = {{Kubernetes Community}},
  title = {Pod Quality of Service Classes},
  howpublished = {Kubernetes Documentation},
  year = {2024},
  note = {\url{https://kubernetes.io/docs/concepts/workloads/pods/pod-qos/}},
}

@misc{systemd-oomd,
  author = {{systemd Project}},
  title = {systemd-oomd.service --- A Userspace Out-Of-Memory (OOM) Killer},
  howpublished = {systemd Documentation},
  year = {2024},
  note = {\url{https://www.freedesktop.org/software/systemd/man/systemd-oomd.service.html}},
}

@misc{psi,
  author = {Weiner, Johannes},
  title = {{PSI} --- Pressure Stall Information},
  howpublished = {Linux Kernel Documentation},
  year = {2018},
  note = {\url{https://docs.kernel.org/accounting/psi.html}},
}

@misc{k8s-inplace-resize,
  author = {{Kubernetes Community}},
  title = {In-Place Update of Pod Resources},
  howpublished = {Kubernetes Enhancement Proposals},
  year = {2024},
  note = {\url{https://github.com/kubernetes/enhancements/tree/master/keps/sig-node/1287-in-place-update-pod-resources}},
}

@misc{meta-oomd,
  author = {{Meta/Facebook}},
  title = {oomd: A Userspace Out-of-Memory Killer},
  howpublished = {GitHub},
  year = {2024},
  note = {\url{https://github.com/facebookincubator/oomd}},
}

@misc{kim2025costdynamicreasoning,
  author = {Kim, Jiin and Shin, Byeongjun and Chung, Jinha and Rhu, Minsoo},
  title = {The Cost of Dynamic Reasoning: Demystifying {AI} Agents and Test-Time Scaling from an {AI} Infrastructure Perspective},
  howpublished = {arXiv preprint arXiv:2506.04301},
  year = {2025},
}

@misc{mei2024aios,
  author = {Mei, Kai and others},
  title = {{AIOS}: {LLM} Agent Operating System},
  howpublished = {arXiv preprint arXiv:2403.16971},
  year = {2024},
}

@misc{agentsight2025,
  author = {Zheng, Yusheng and Hu, Yanpeng and Yu, Tong and Quinn, Andi},
  title = {{AgentSight}: System-Level Observability for {AI} Agents Using {eBPF}},
  howpublished = {arXiv preprint arXiv:2508.02736},
  year = {2025},
}

@article{wang_llm_agent_survey,
  author  = {Wang, Lei and Ma, Chen and Feng, Xueyang and Zhang, Zeyu and Yang, Hao and
             Zhang, Jingsen and Chen, Zhiyuan and Tang, Jiakai and Chen, Xu and Lin, Yankai and
             Zhao, Wayne Xin and Wei, Zhewei and Wen, Ji-Rong},
  title   = {A Survey on Large Language Model Based Autonomous Agents},
  journal = {Frontiers of Computer Science},
  volume  = {18},
  number  = {6},
  pages   = {186345},
  year    = {2024},
  doi     = {10.1007/s11704-024-40231-1},
}

@inproceedings{kgent,
  author    = {Zheng, Yusheng and Yang, Yiwei and Chen, Maolin and Quinn, Andrew},
  title     = {{Kgent}: Kernel Extensions Large Language Model Agent},
  booktitle = {Proceedings of the ACM SIGCOMM 2024 Workshop on eBPF and Kernel Extensions},
  series    = {eBPF '24},
  year      = {2024},
  pages     = {30--36},
  publisher = {ACM},
  doi       = {10.1145/3672197.3673434},
}

@inproceedings{react,
  author = {Yao, Shunyu and Zhao, Jeffrey and Yu, Dian and Du, Nan and Shafran, Izhak and Narasimhan, Karthik and Cao, Yuan},
  title = {{ReAct}: Synergizing Reasoning and Acting in Language Models},
  booktitle = {International Conference on Learning Representations (ICLR)},
  year = {2023},
}

@misc{bodea2025trusted,
  author        = {Bodea, Teofil and Misono, Masanori and Pritzi, Julian and Sabanic, Patrick and Sommer, Thore and Unnibhavi, Harshavardhan and Schall, David and Santos, Nuno and Stavrakakis, Dimitrios and Bhatotia, Pramod},
  title         = {Trusted {AI} Agents in the Cloud},
  howpublished  = {arXiv preprint arXiv:2512.05951},
  year          = {2025},
}

@misc{chen2025kairos,
  author        = {Chen, Jinyuan and Shi, Jiuchen and Chen, Quan and Guo, Minyi},
  title         = {Kairos: Low-latency Multi-Agent Serving with Shared {LLM}s and Excessive Loads in the Public Cloud},
  howpublished  = {arXiv preprint arXiv:2508.06948},
  year          = {2025},
}

@misc{ni2025astraea,
  author        = {Ni, Hongqiu and Zhang, Jiabao and Li, Guopeng and Wang, Zilong and Wu, Ruiqi and Zhang, Chi and Tan, Haisheng},
  title         = {Astraea: A State-Aware Scheduling Engine for {LLM}-Powered Agents},
  howpublished  = {arXiv preprint arXiv:2512.14142},
  year          = {2025},
}

@misc{li2025continuum,
  author        = {Li, Hanchen and Mang, Qiuyang and He, Runyuan and Zhang, Qizheng and Mao, Huanzhi and Chen, Xiaokun and Zhou, Hangrui and Cheung, Alvin and Gonzalez, Joseph and Stoica, Ion},
  title         = {Continuum: Efficient and Robust Multi-Turn {LLM} Agent Scheduling with {KV} Cache Time-to-Live},
  howpublished  = {arXiv preprint arXiv:2511.02230},
  year          = {2025},
}

@misc{asgar2025heterogeneous,
  author        = {Asgar, Zain and Nguyen, Michelle and Katti, Sachin},
  title         = {Efficient and Scalable Agentic {AI} with Heterogeneous Systems},
  howpublished  = {arXiv preprint arXiv:2507.19635},
  year          = {2025},
}

@misc{claude-secure-deployment,
  author       = {{Anthropic}},
  title        = {Securely Deploying {AI} Agents},
  howpublished = {Claude API Documentation (Agent SDK Guides)},
  year         = {2026},
  note         = {\url{https://platform.claude.com/docs/en/agent-sdk/secure-deployment} (accessed 2026-02-09)},
}

@misc{claude-code-sandboxing,
  author       = {Dworken, David and Weller-Davies, Oliver},
  title        = {Beyond Permission Prompts: Making {Claude Code} More Secure and Autonomous},
  howpublished = {Anthropic Engineering Blog},
  year         = {2025},
  note         = {Published Oct 20, 2025. \url{https://www.anthropic.com/engineering/claude-code-sandboxing} (accessed 2026-02-09)},
}

@misc{openai-codex,
  author       = {{OpenAI}},
  title        = {Introducing Codex},
  howpublished = {OpenAI},
  year         = {2025},
  month        = may,
  note         = {\url{https://openai.com/index/introducing-codex/} (accessed 2026-02-09)},
}

@misc{github-copilot-coding-agent,
  author       = {{GitHub}},
  title        = {About {GitHub} Copilot coding agent},
  howpublished = {GitHub Documentation},
  year         = {2026},
  note         = {\url{https://docs.github.com/en/copilot/concepts/agents/coding-agent/about-coding-agent} (accessed 2026-02-09)},
}

@misc{google-jules,
  author       = {{Google}},
  title        = {Jules, {Google}'s asynchronous {AI} coding agent, is out of public beta},
  howpublished = {Google Blog},
  year         = {2025},
  month        = aug,
  note         = {\url{https://blog.google/innovation-and-ai/models-and-research/google-labs/jules-now-available/} (accessed 2026-02-09)},
}

@misc{cognition-devin,
  author       = {{Cognition}},
  title        = {Introducing Devin, the first {AI} software engineer},
  howpublished = {Cognition Blog},
  year         = {2024},
  month        = mar,
  note         = {\url{https://cognition.ai/blog/introducing-devin} (accessed 2026-02-09)},
}

@misc{apple-xcode-agentic-coding,
  author       = {{Apple}},
  title        = {Xcode 26.3 unlocks the power of agentic coding},
  howpublished = {Apple Newsroom},
  year         = {2026},
  month        = feb,
  note         = {\url{https://www.apple.com/newsroom/2026/02/xcode-26-point-3-unlocks-the-power-of-agentic-coding/} (accessed 2026-02-09)},
}

@misc{cursor-agent,
  author       = {{Cursor Team}},
  title        = {Best practices for coding with agents},
  howpublished = {Cursor Blog},
  year         = {2026},
  month        = jan,
  note         = {\url{https://cursor.com/blog/agent-best-practices} (accessed 2026-02-09)},
}

@inproceedings{swe-bench,
  author = {Jimenez, Carlos E. and Yang, John and Wettig, Alexander and Yao, Shunyu and Pei, Kexin and Press, Ofir and Narasimhan, Karthik R.},
  title = {{SWE-bench}: Can Language Models Resolve Real-World {GitHub} Issues?},
  booktitle = {International Conference on Learning Representations (ICLR)},
  year = {2024},
  note = {Oral. arXiv:2310.06770},
}

\end{document}